\documentclass[aps,prl,twocolumn,showpacs,floatfix,a4paper]{revtex4}
\usepackage{graphicx}
\usepackage{amsmath}
\usepackage{amssymb}
\usepackage{bm}
\usepackage[usenames]{color}
\usepackage{amsfonts}
\usepackage{appendix}
\usepackage[normalem]{ulem}

\graphicspath{ {images/} }
\let\oldsqrt\sqrt
\def\sqrt{\mathpalette\DHLhksqrt}
\def\DHLhksqrt#1#2{\setbox0=\hbox{$#1\oldsqrt{#2\,}$}\dimen0=\ht0
\advance\dimen0-0.2\ht0
\setbox2=\hbox{\vrule height\ht0 depth -\dimen0}{\box0\lower0.4pt\box2}}
\DeclareMathSizes{10}{10}{6}{5}

 \newcommand{\be}{\begin{equation} }
 \newcommand{\ee}{\end{equation} }
  \newcommand{\ba}{\begin{aligned} }
  \newcommand{\ea}{\end{aligned} }

\begin{document}

\title{Entanglement entropy distribution in the strongly disordered 
one-dimensional Anderson model}
\author{B. Friedman}
\author{R. Berkovits}
\affiliation{Department of Physics, Jack and Pearl Resnick Institute, Bar-Ilan University, Ramat-Gan 52900, Israel}

\begin{abstract}
  The entanglement entropy
  distribution of strongly disordered one dimensional spin chains, which
  are equivalent to spinless fermions at half-filling on a bond (hopping) disordered
  one-dimensional Anderson model, has been shown to  exhibit very distinct
  features such as peaks at integer multiplications of $\ln(2)$, essentially
  counting the number of singlets traversing the boundary. Here we show that
  for a canonical Anderson model with box distribution
  on-site disorder and  repulsive
  nearest-neighbor interactions the entanglement entropy distribution
  also exhibits interesting features,  albeit different than the distribution
  seen for the bond disordered Anderson model. The canonical Anderson model
  shows a broad peak at low entanglement values and one narrower peak at $\ln(2)$.
  Density matrix renormalization group (DMRG) calculations reveal this structure and the influence of the
  disorder strength and the interaction strength on its shape. A modified 
 real space renormalization group (RSRG) method was used to get a better understanding of this behavior. As
  might be expected the peak centered at low values of entanglement entropy 
  has a tendency to shift to lower values as disorder is enhanced.
  A second peak appears around the entanglement entropy value of $\ln(2)$, 
  this peak is broadened and no additional peaks at higher integer 
  multiplications of $\ln(2)$ are seen. We attribute the differences
  in the distribution between the canonical model and the broad hopping disorder
  to the influence of the on-site disorder which breaks the symmetry 
  across the boundary.

\end{abstract}

\pacs{72.15.Rn, 73.20.Fz, 73.21.Hb}
\maketitle

\section{Introduction}

 There has been much recent interest in the distribution of the entanglement 
entropy (EE) for strongly disordered 1D systems 
\cite{laf,ramiraz,berkovits12,berkovits15}. For a system in a pure state 
$|\psi\rangle$, the EE is given by $S=-\sum_i\mu_i\log(\mu_i)$, with 
$\mu_i$ eigenvalues of $\rho_{A}=Tr_{B}|\psi\rangle\langle\psi|$, the reduced 
density matrix (RDM) over a sub-region of length $L_A$ of the system, 
where the degrees of freedom of the remaining area $B$ are traced out.  
Quantities such as the mean or the median cannot
fully capture the behavior of a disordered system which is mesoscopic
in nature \cite{imry02,akkermans07}.
It is therefore essential to describe the EE using its distribution, rising 
from different disorder realization.

Most of the conclusions regarding the EE distribution were obtained for 
spin chains with a power law distribution of nearest-neighbor coupling
between the spins 
and no magnetic field\cite{laf,ramiraz}.
This facilitates the use of the real-space renormalization group (RSRG) method
\cite{das,fisher,moore} which enables the treatment of large systems. 
The main finding of these studies is that there is a distinct distribution
characterized by peaks at integer multiplies of $\ln(2)$. The details
depend on boundary conditions and the number of spins in the entangled region
(even/odd). This behavior stems from 
spin singlets straddling the boundary between the regions, corresponding in
the fermionic language to an electron resonating between locations across
the boundary.

The above mentioned spin model translates into a fermionic Anderson model with
bond (i.e., hopping) disorder, but no on-site disorder.
Spin singlets straddling the boundary translates
to an electron resonating between locations across
the boundary for the Anderson model.
Nevertheless, we must be cautious about applying these conclusions to
generic cases of disordered 1D systems, since the bond disordered Anderson
model has peculiarities
such as a divergence in the density of states at the middle
of the band \cite{dyson53} accompanied by the appearance of extended
states \cite{theodorou76,eggarter78}.
The authors of ref.\cite{li} studied the highly excited states
  of a Heisenberg spin $1/2$ chain with random magnetic field and found
  a peak around zero and a peak very close to $\ln(2)$. This result hints that
  the on-site term has a crucial influence on the form of the EE distribution.
  However, these authors discuss high energy states in a small system. As is
  well known the EE of the ground state behaves differently than the EE of
  excited states (area law vs. volume law). Thus,
  further research is needed to see the behavior of the ground state EE 
distribution.
  
Indeed, in this paper we would like to see whether the 
behavior of the EE distribution depends on the disorder, i.e., whether there
is a difference between hopping and on-site disorder.
We therefore study the canonical Anderson model in the presence of a box 
distribution on-site disorder. We add also nearest neighbor electron-electron 
interactions for two reasons: The first is to clarify whether the interactions
which are known to increase the effect of disorder (i.e. to shorten the 
localization length \cite{apel_82,giam,berkovits12a}) 
influence the EE distribution differently than the 
on-site disorder. The second stems from using the density matrix 
renormalization group (DMRG) method. Adding interaction is a way to enhance
disorder without changing the on-site energy distribution width. Since while 
using DMRG the accuracy
degrades quite rapidly as the width of the on-site disorder distribution grows,
increasing the interaction is a viable way to
increase the effective disorder.

The following results may be garnered from the DMRG calculations:
(i) The distribution exhibits {\it one} peak at the value of $\ln(2)$, but
no peaks at higher integer multiplication of $\ln(2)$.
(ii) There is an additional broad skewed peak at lower values of the EE.
This peak shifts
to lower values and becomes more skewed as the disorder grows.
(iii) The numerical data indicates that the distribution does not scale 
exclusively by the
localization length $\xi$ (which for the Anderson model is a function
of the interaction $U$ and the width of the distribution of on-site
disorder $W$\cite{apel_82,giam,berkovits12a}).

Generally, as the on-site disorder or interaction increases the EE,
distribution seems both to shift the location of the main peak to lower values
of EE (as would have been naively expected) and develop a second, lower peak 
at $\ln(2)$. The second peak is reminiscent of the behavior of the
power-law distributed
bond model \cite{laf,ramiraz}, although no additional peaks
at higher multiples of  $\ln(2)$ are seen.
The similarity between
  these results and the conclusions of ref. \cite{li} indicate
  that this is a rather universal property of generic disorder.
We introduce a modified RSRG method that incorporates on-site disorder.
The numerical renormalization procedure shows 
a main peak at very low EE and an additional peak around $\ln(2)$, thus 
capturing the main features of the DMRG calculation.

\section{The model}

We consider a spinless fermions system at half-filling with on-site 
disorder and a repulsive nearest-neighbor interaction:

\be
\ba
H=&\sum_{j=1}^{L} \epsilon_j c_j^{\dagger} c_j-t\sum_{j=1}^{L-1} 
c_{j}^{\dagger}c_{j+1}+h.c.\\
&+U\sum_{j=1}^{L-1}(c_j^{\dagger} c_j-\frac{1}{2}) 
(c_{j+1}^{\dagger} c_{j+1}-\frac{1}{2})
\label{h}
\ea
\ee
where $c_j$ is the vacuum annihilation operator. The on-site disorder term $\epsilon_j$ is chosen to distribute uniformly in the range $[-W/2,W/2]$.

\section{DMRG results}

The DMRG \cite{white,scholl} is a very accurate numerical method
for calculating the ground state of the disordered interacting
1D system and for the calculation of the reduced density
matrix \cite{berkovits15,berkovits12a}. Here we consider a system of length 
$L=700$, and strength of disorder $W=0.7,1.5,2.5,3,3.5,4,5$. These strengths' 
of disorder corresponds to $\xi \sim 214.3, 46.7,16.8,11.7,8.6,6.6,4.2$ correspondingly
for the non-interacting case. 
For finite samples the quantum phase is defined by the relation between
the correlation (localization) length $\xi$ and the size of the sample $L$.
The regime for which $L \gg \xi$ which is 
usually considered as strongly disorder is the subject of the current paper,
and occurs for example at $W=2.5,3,3.5,4,5$. Other
values of disorder for which $\xi\sim L$ (effectively a metallic phase)
and for attractive nearest-neighbor interaction
which result in a 
superconducting phase are widely discussed in ref. \cite{berkovits15}.

For the interacting case, 
using renormalization group \cite{apel_82} the localization length 
dependence on interaction strength can be formulated as
$\xi(W,U) = (\xi(W,U=0))^{1/(3-2g(U))}$,
where $g(U)=\pi/[2 \cos ^{-1} (-U/2)]$ is the Luttinger 
parameter \cite{g_formula}. 
For non-interacting electrons $g(U=0)=1$.
Since for repulsive interactions $g<1$ decreases as a 
function of the interaction strength, one finds that the localization length 
always decreases as a function of the repulsive interaction strength.


The distribution of the EE for different
values of disorder $W$ and interaction $U$ are calculated.
For each of the different realizations of disorder the EE
is calculated for different lengths of the region A, 
$L_A=L/4,L/4+10,L/4+20,\ldots,3L/4$. Since the distribution of the
EE is very similar for different values of $L_A$ as long as $L_A$
is not too close to the edge \cite{berkovits15}, 
we accumulate statistics on the distribution of
the EE, $P(s)$, 
for $100$ realizations of disorder at any given disorder and interaction
strength for the different values of $L_A$.

\begin{figure}[ht!]
\centering
\includegraphics[width=85mm]{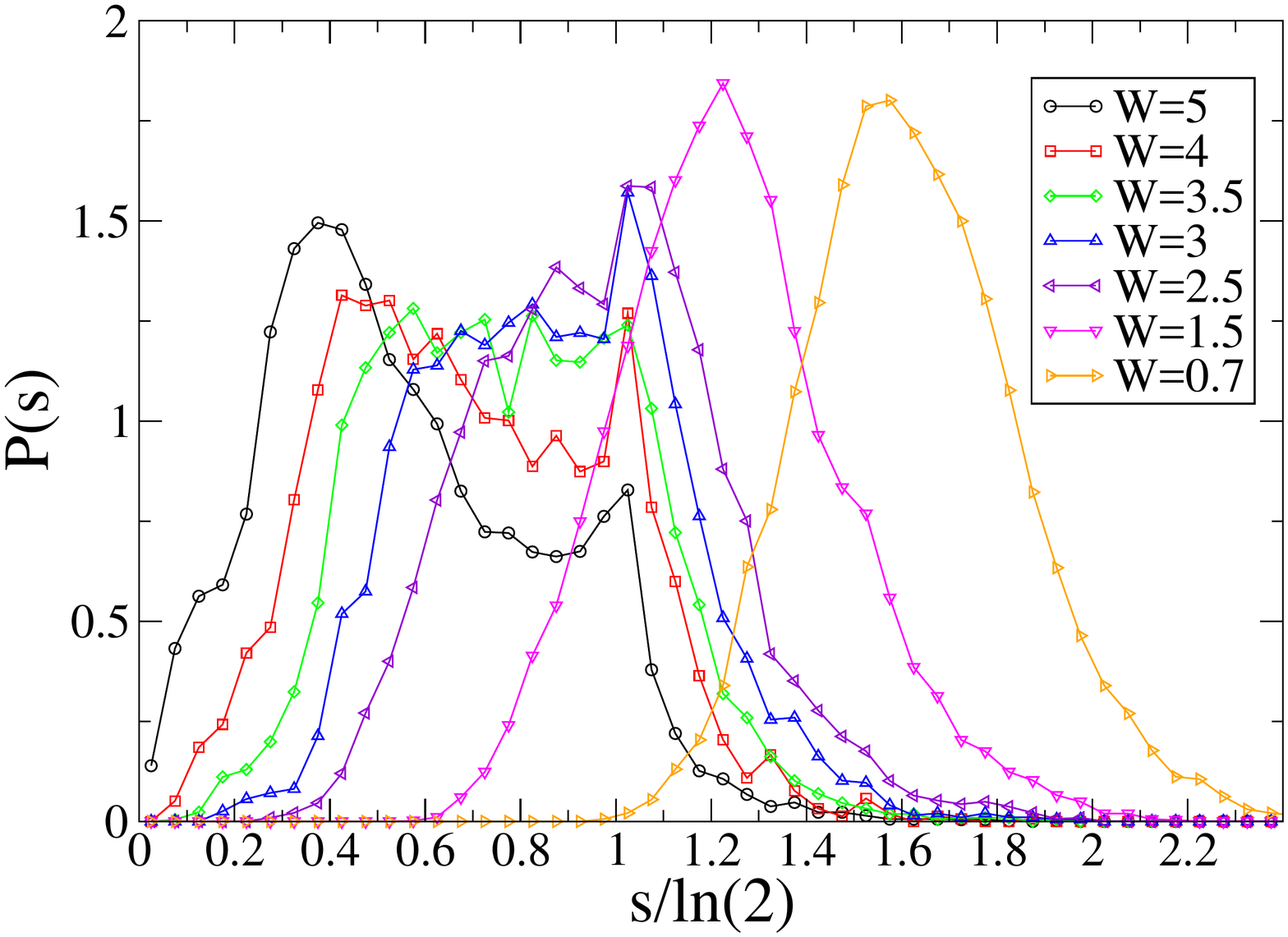}
\caption{DMRG results for the EE distribution for different strength
of disorder $W$ for a sample length $L=700$ and no electron-electron
interaction ($U=0$). As the on-site disorder, $W$,
becomes stronger, the center of the distribution moves to smaller values of $s$.
A secondary peak develops at $\ln(2)$ and the distribution becomes more 
skewed to the right.}
\label{fig1}

\end{figure}

\begin{figure}[ht!]
\centering
\includegraphics[width=85mm]{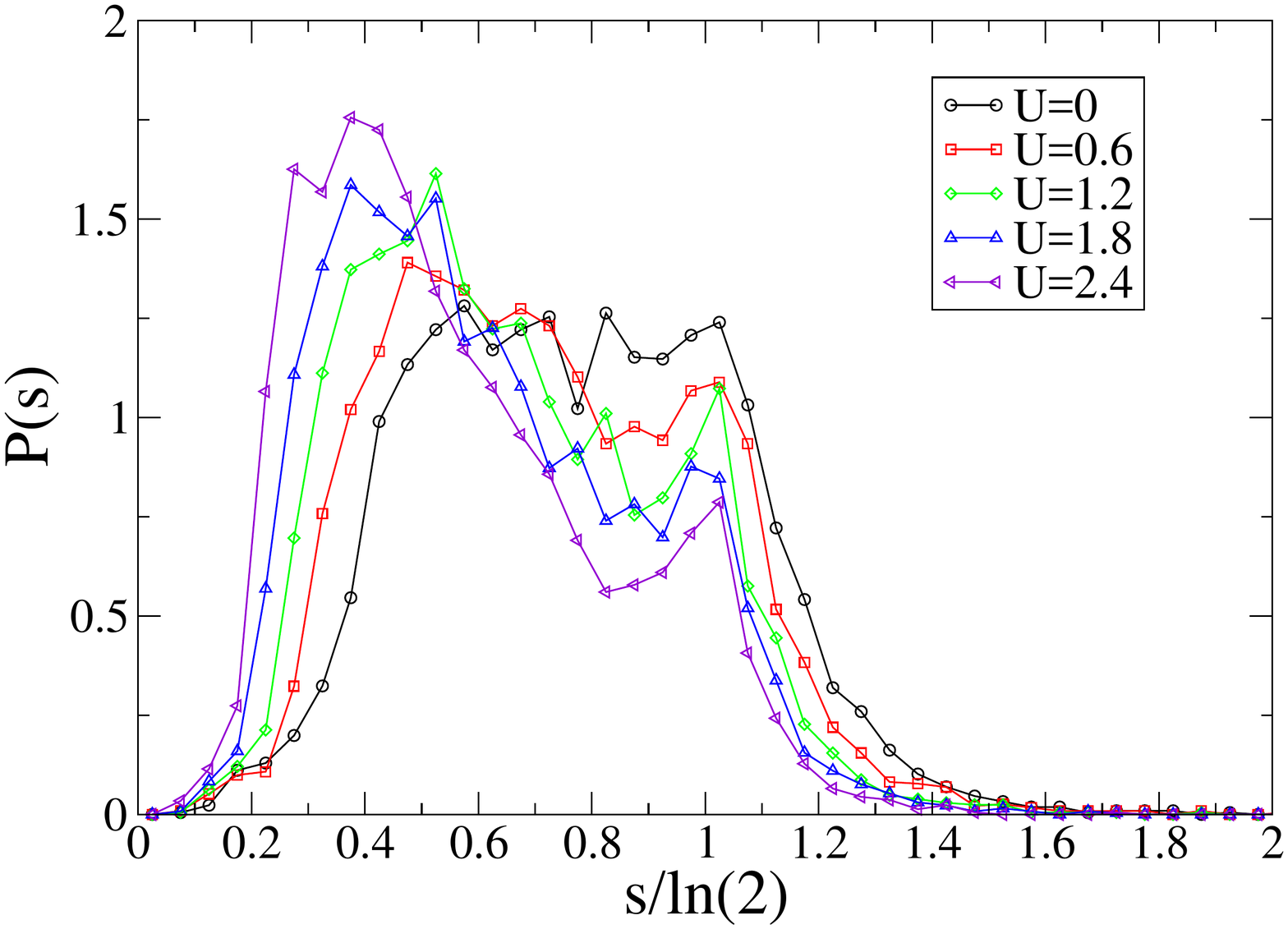}
\caption{DMRG results for the EE distribution for different strength
of electron-electron interactions $U$ for a sample length $L=700$, and fixed 
on-site disorder, $W=3.5$. As the interactions
become stronger, a similar behavior to the one observed for an increase in
$W$ is seen (see Fig. \ref{fig1}).}
\label{fig2}
\end{figure}

\begin{figure}[ht!]
\centering
\includegraphics[width=85mm]{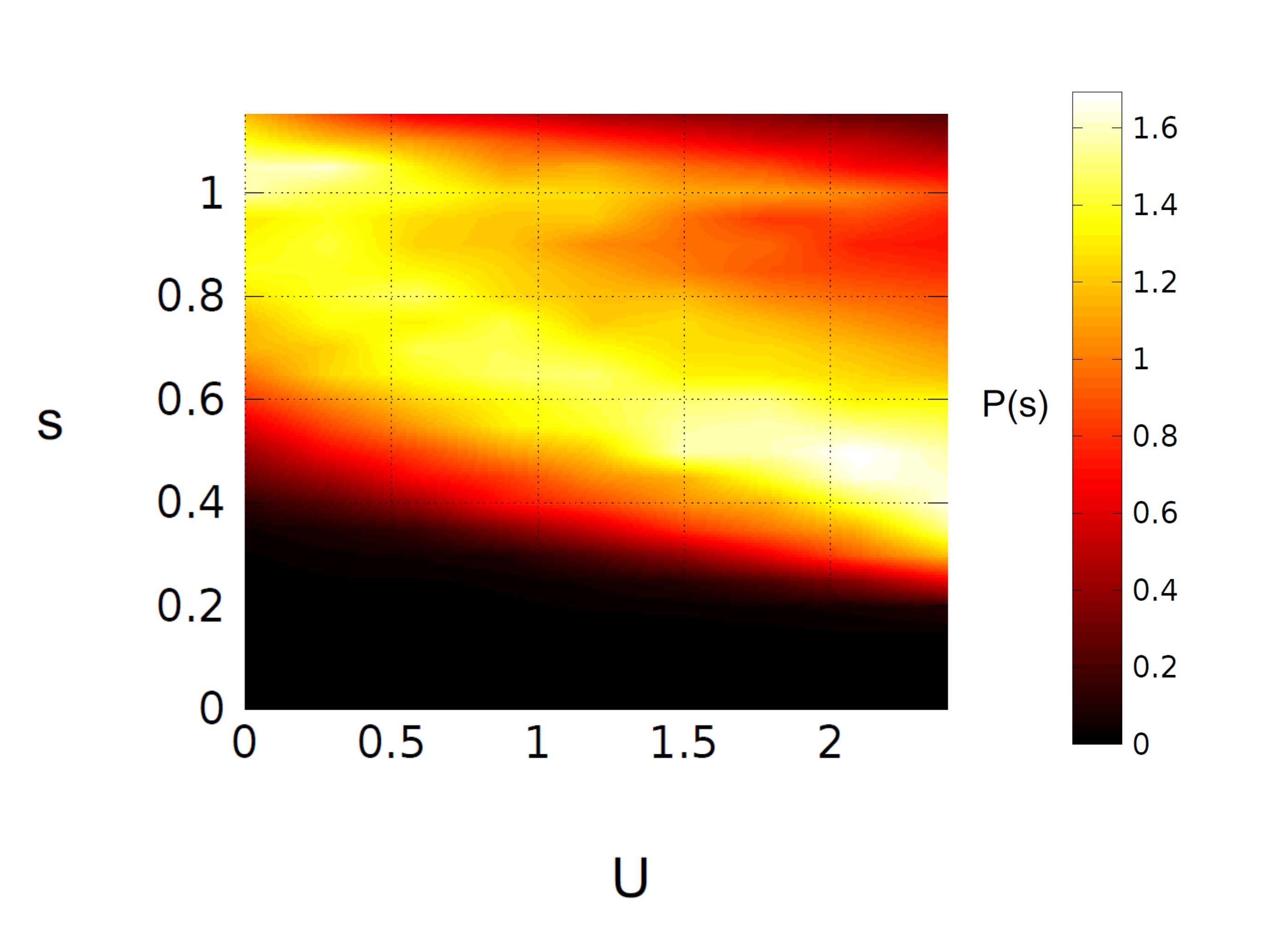}
\caption{A color map of the distribution of the EE as function
of the interaction strength for $L=700$, and $W=2.5$. The main peak shifts 
to lower values of $s$
and becomes more skewed, while a secondary peak is pegged to $s=\ln(2)$. }
\label{fig3}
\end{figure}

In Fig. \ref{fig1} the distribution of the EE for different width of the box 
distribution $W$ with no interactions ($U=0$) is plotted. For weak 
disorder ($W=0.7$), the distribution is Gaussian \cite{berkovits15},
as disorder
increases the distribution center moves to lower values of the disorder,
becomes skewed to the left, and develops a second peak at $s=\ln(2)$. 
As can be seen in Fig. \ref{fig2}, an 
essentially similar behavior is seen when we keep the disorder
fixed (at $W=3.5$) but change $U$. This behavior is accentuated in Fig.
\ref{fig3}, where a color map is presented for a different value
of on-site disorder ($W=2.5$). The shift of the main peak to lower
values of $s$, as well as the emergence of a second smaller peak at
$s=\ln(2)$ is evident.


How can we understand this behavior?
Let us first recall the results for the extreme disorder discussed for a spin
chain with a power-law distribution of the $J$ coupling (i.e., 
infinite variance), which maps onto the spinless fermionic 
model of Eq. (\ref{h}) with a power-law distribution of the hopping element $t$
and no on-site disorder. Using RSRG it has been shown that for the 
periodic  boundary conditions and an even length of $L_A$ there are peaks in P(s)
at values of $s=2n\ln(2)$, where $n$ is an integer \cite{laf,moore}. For hard
wall boundary conditions there are peaks in the distribution for 
$s=n\ln(2)$. These peaks stem from singlet states between spins straddling the 
boundary between regions, which for the fermion version corresponds to a
resonating electron between two sites across the boundary. Each such
bond crossing the boundary between region A and B results in a contribution
to the EE of $\ln(2)$. For the extreme disorder case these peaks dominate 
the distribution.

\begin{figure}[ht!]
\centering
\includegraphics[width=85mm]{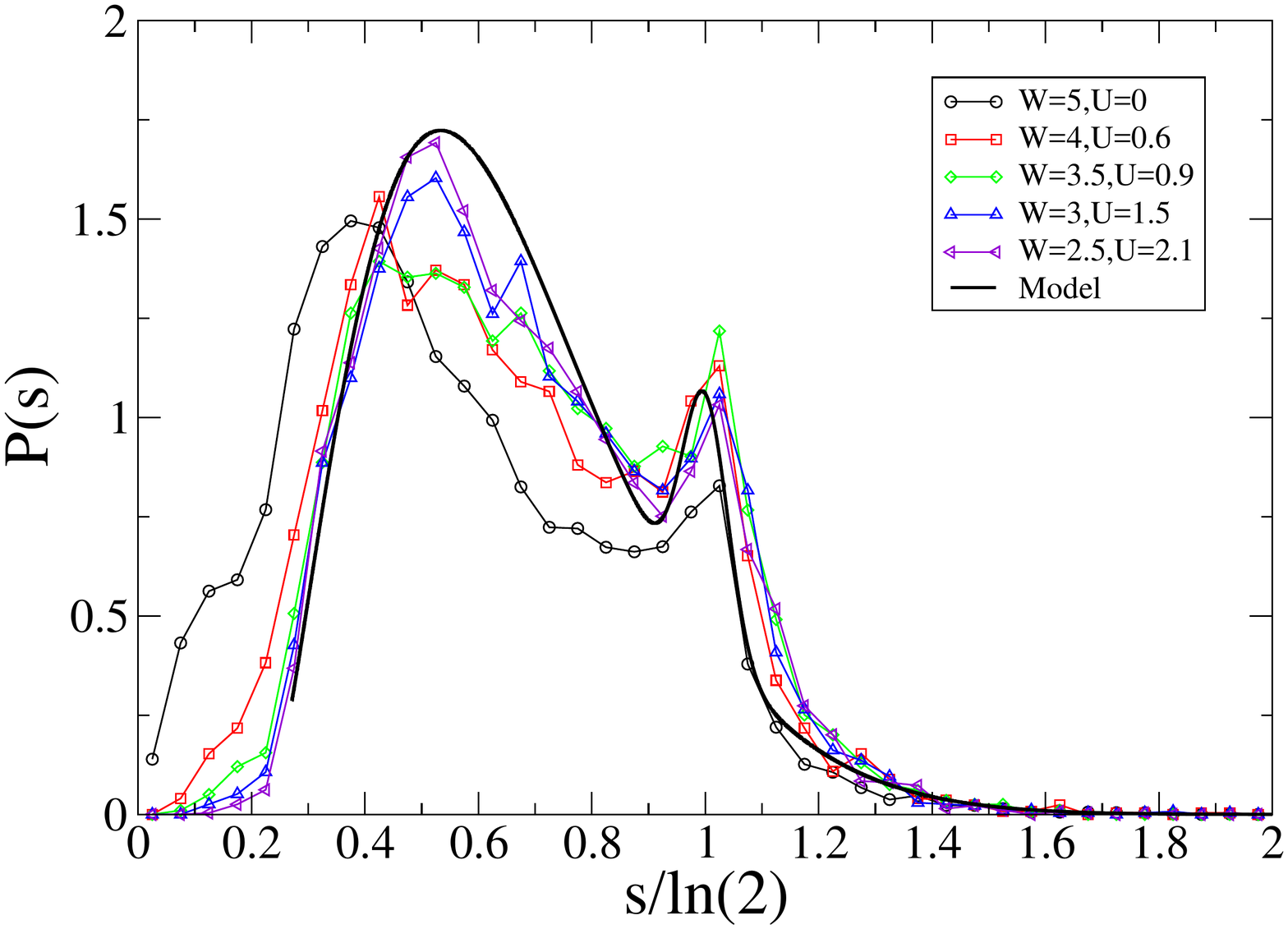}
\caption{The distribution of $P(s)$ for different values of $W$ and $U$
which correspond to the same value $\xi \sim 4.2$ for a $L=700$ system.
Although the curves do not exactly scale, nevertheless for higher values
of interaction they more or less fall on top of each other. }
\label{fig4}
\end{figure}

This is not the case here. As disorder or interaction increases, a low EE
peak ($s \ll \ln(2)$) becomes dominant, while no higher peaks beyond the
$s=\ln(2)$ appear. One may also wonder whether the distribution
curve depends on the localization length $\xi$ exclusively, i.e.,
can the distribution of the EE be scaled by $P(\xi(U,W))$?
This question is addressed in Fig. \ref{fig4}, where the distribution
of systems with the same localization length $\xi$ but different
values of $W$ and $U$ are considered. Although the curves are generally similar,
there is no perfect scaling between the curves. Nevertheless, the general
features of the curves are the same, e.g., a skewed two peaked distribution,
with a broad peak at small values of $s$ and a narrow peak at $\ln(2)$.

In order to find an analytic description for the DMRG results we simplify our system to the following toy model: 
We describe a toy model containing four sites and two fermions with nearest-neighbors interaction $U$ and a defect, which is the source of disorder in the model (Fig. \ref{model}). The presence of the defect adjusts the value of the hopping parameter $t$ across the defect. The modified hopping parameter is denoted $\tilde{t}$.
By choosing different values of $\tilde{t}$, we create an ensemble of toy-model systems, and examine its statistics.  
The position of the impurity naturally divides the system into two (identical) sub-systems denoted A and B.
The system is described by the following Hamiltonian:
\begin{equation}
\begin{aligned}
H=&t (a_1^{\dagger}a_{2}+b_3^{\dagger}b_{4})+h.c.+\tilde{t}(a_2^{\dagger}b_3+b_3^{\dagger}a_2)\\
&-U(n_{a_1}n_{a_2}+n_{a_2}n_{b_3}+n_{b_3}n_{b_4})
\end{aligned}
\label{toyh}
\end{equation}

$a_i(b_i)$ is the annihilation operator of site $i$ in region $A(B)$ and $n_{c_k}=c_k^{\dagger}c_k$ is the fermionic number operator. In this model a minus sign appears in front of the interaction strength $U$, therefore a repulsive interaction is obtained by negative values of $U$.  

The size of the model allows an analytic calculation of the entanglement entropy. However, the full result is too long to present and use. Hence, we use Taylor expansion and present the first few terms, which give a good approximation:
\be
\ba
&S_A\approx -\frac{\sqrt{3}}{4} \cosh^{-1}(7) + \log(4) -
 \frac{(-6 + \tilde{t}^2) \cosh^{-1}(7)}{8 \sqrt{6} U}+\\
&\frac{1}{576 U^2} (\sqrt{3} (276 + 12 \tilde{t}^2 + \tilde{t}^4)\cosh^{-1}(7)+  \\
   &-12 (36 - 36 \tilde{t}^2 + \tilde{t}^4 + 48 \tilde{t}^2 \log(|\frac{\tilde{t}}{U}|)) +O(U^{-3})\\
   \ea
\ee

As detailed in appendix A, when
assuming a Gaussian distribution for $\tilde{t}$, 
we can find the entanglement distribution analytically.
The toy model captures the physics of the low-entanglement peak
successfully, but fails to describe the second peak, probably due to its size.
We can try to fit the distribution obtained by DMRG by
a heuristic  description plotted in 
Fig. \ref{fig4},
where the distribution 
is the sum of a broad Gaussian centered at the lower peak 
position and a narrow Gaussian centered at $\ln(2)$. Hence,

\be
\ba
&P(s)=\frac{D_1}{\sqrt{2\pi}\sigma_1}e^{-\dfrac{(s-\mu_1)^2}{2\sigma_1^2}}
+\frac{D_2}{\sqrt{2\pi}\sigma_2} e^{-\dfrac{(s-\mu_2)^2}{2\sigma_2^2}}.
\ea
\label{analyticP}
\ee
Here $\mu_1\propto \ln(erf(L_A/\xi))+const $ \cite{berkovits12a},
$\mu_2\approx \ln(2), D_1\propto \dfrac{\partial \tilde{t}}{\partial s}$
(see appendix A for expression),  $\sigma_{1,2}$ inversely depend on $W,U$
and $D_{2}$ is a constant. The explicit numerical values of 
these constants are determined from a fit.
This seems to indicate that there are two distinct processes
contributing to the distribution. One centered at $\ln(2)$ and the second
around low values of $s$. This will be discussed some more
at the end of the next section.

\section{Modified real space renormalization group}
\begin{figure}[ht!]
\centering
\includegraphics[width=85mm]{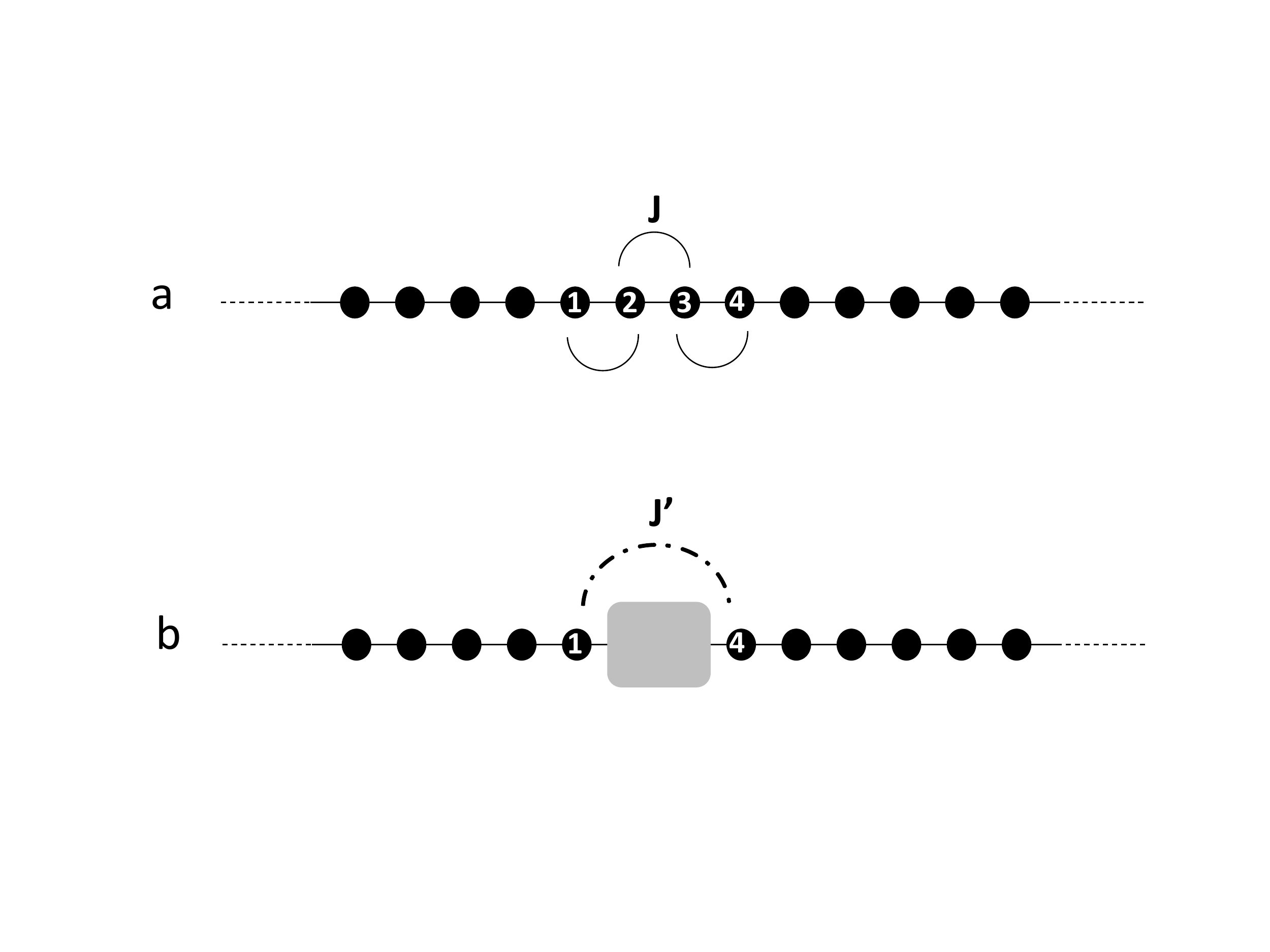}
\caption{Real space renormalization group (RSRG) scheme:
a) One detects the strongest bond J on the chain and assumes a singlet state for the two sites connected by J. Then one calculates the effect of this singlet on the nearest neighbors from both sides using perturbation theory. The result of the calculation is an effective coupling between the neighbors, J'.
b) One eliminates the sites connected by J and rewrite the chain with the effective coupling J'.}
\label{rsrg}
\end{figure}

 Using the canonical Jordan-Wigner transformation on the 
Hamiltonian (\ref{h}) the Anderson model may be rewritten as the XXZ spin model 
in the presence of a random magnetic field:
 \begin{multline}
 H=\sum_{j=1}^{N-1}[ -2t(S_j^xS_{j+1}^x+ S_j^yS_{j+1}^y)\\+\epsilon_jS_j^z+U S_j^zS_{j+1}^z]+\epsilon_{N}S^z_{N},
 \label{hs}
 \end{multline}
where $S_j^i$ is the spin operator of the $j$-th spin in the $i=x,y,z$
direction
and we ignore an overall constant energy term.
Similar models (albeit with no random magnetic term
nor interaction term - the second line in Eq. (\ref{hs})) 
have been previously studied using the RSRG 
\cite{das,fisher,moore} method.
RSRG  assumes that two neighboring spins coupled by
the strongest bond in the system are in their
ground state. Calculating the effect of this state on the two
neighboring spins using second order perturbation theory and rewriting the Hamiltonian accordingly, 
 all three bonds are replaced with an effective bond between the neighbors (see Fig. \ref{rsrg}). 
Iterating over the system, one can find an approximate ground state. 
For example in the random Heisenberg xx-model, the ground state of a single 
bond is a singlet. Fig. \ref{rsrg} shows the renormalization process for this spin model.

For the non-interacting fermion system which is equivalent to the 
Heisenberg xx model the ground state turns out to be {in} the 
"random singlet phase" (RSP) which contains pairs of arbitrarily 
long singlets over the whole system.

Using this renormalization procedure the
EE  may be calculated and it turns out that the EE is proportional to the 
number of singlets pairs straddling the boundary between 
region A and B \cite{moore}.
Thus, the EE distribution exhibits multiple peaks \cite{laf,ramiraz} 
corresponding to the value of the EE of a singlet ($\ln(2)$) multiplied
by the even (odd) number of singlets.
Whether one has even (odd) number of singlets is related to even (odd)  
number of sites in region A.

When one considers also on-site disorder and nearest neighbor interactions
(Eq.(\ref{hs})) there isn't a single parameter that determines 
the ground state of  two coupled spins, since the energy levels of a single 
bond depend on four different parameters. Thus, the classical RSRG is not 
applicable to this system. Instead we modify the RSRG scheme to
take the additional parameters into account.

Let us explicitly write
Hamiltonian (\ref{hs}) reduced to sites $2$ and $3$ shown in Fig.\ref{rsrg}.
In the basis $\{ |\uparrow\uparrow\rangle, |\downarrow\downarrow\rangle, |\uparrow\downarrow\rangle, |\downarrow\uparrow\rangle\}$
this takes the form:
\be
H^{2-3} = \left( \begin{array}{cccc}
|\uparrow\uparrow\rangle &|\downarrow\downarrow\rangle & |\uparrow\downarrow\rangle &|\downarrow\uparrow\rangle   \\
E_1 & 0 & 0 & 0   \\
0& E_2 & 0 & 0   \\
0 & 0 & -\frac{U_2}{4}+\frac{(\epsilon_2-\epsilon_3)}{2} & -t_2 \\
0 & 0 & -t_2 & -\frac{U_2}{4}+\frac{(\epsilon_3-\epsilon_2)}{2} \\
\end{array} \right).
\label{singlebond}
\ee
Diagonalizing the matrix we get the following eigenvectors

\be
\ba
&|v_1\rangle=|\uparrow\uparrow\rangle\\
&|v_2\rangle=|\downarrow\downarrow\rangle\\
&|v_3\rangle=a_3|\uparrow\downarrow\rangle +b_3|\downarrow\uparrow\rangle \\
&|v_4\rangle=a_4|\uparrow\downarrow\rangle +b_4|\downarrow\uparrow\rangle \\
\ea
\ee
where 
\be
\ba
&a_{3,4}=\dfrac{\epsilon_3-\epsilon_2\pm\sqrt{(\epsilon_3-\epsilon_2)^2+4t_2^2}}{2t_2N_{3,4}},\\
& b_{3,4}=\dfrac{1}{N_{3,4}},\\
& N_{3,4}=\sqrt{(\dfrac{\epsilon_3-\epsilon_2\pm\sqrt{(\epsilon_3-\epsilon_2)^2+4t_2^2}}{2t_2})^2+1}.
\ea
\ee
and eigenenergies:
\be
\begin{aligned}
E_1=U_2/4+(\epsilon_2+\epsilon_3)/2\\
E_2=U_2/4-(\epsilon_2+\epsilon_3)/2\\
E_3=(-U_2-2\sqrt{(\epsilon_2-\epsilon_3)^2+4t_2^2})/4\\
E_4=(-U_2+2\sqrt{(\epsilon_2-\epsilon_3)^2+4t_2^2})/4\\.
\end{aligned}
\label{energylevels}
\ee
$E_4$ is always larger than $E_3$ and therefore cannot be the ground state. 
The remaining states may be the ground state, depending on the 
parameters. 
We explicitly calculate the energies $E_i,i=1,2,3$, for each bond in the system and find its' ground state. 
If the ground-state of the bond is
$|v_3\rangle$ the 
entanglement entropy across this bond is 
\be
S_{SB}=-|a_3|^2\ln(|a_3|^2)-|b_3|^2\ln(|b_3|^2),
\ee 
while the other possible ground states, $|v_1\rangle$ and $|v_2\rangle$ 
are product states and therefore don't contribute to the EE ($S_{SB}=0$)
Accumulating $S_{SB}$ during the renormalization process will give us the
approximate EE of the system.
We now turn to the renormalization procedure and detect the bond with the 
minimal ground-state energy. We calculate $S_{SB}$ and
assume that second order perturbation theory 
holds. Then, we calculate the effect of the bond ground state on the
neighboring spins. 
 Next, we eliminate the original bonds, and calculate $E_{i}, i=1..3$
for the new  effective bond between spins $1$ and $4$. 

The following Hamiltonian describes the four sites involved in the
renormalization of a single bond (see Fig. \ref{rsrg})
and contains only the terms that are affected in a non trivial
way by the diminished sites.
\be
\ba
H^{1-4}&=-2t_1(S_1^+S_2^-+S_1^-S_2^+)+U_1S_1^zS_2^z\\
&-2t_3(S_3^+S_4^-+S_3^-S_4^+)+U_3S_3^zS_4^z\\
\label{h1234}
\ea
\ee
where $S_j^{\pm}=\dfrac{S_j^x\pm iS_j^y}{2}, j=1..4$.
Indices appear also on all the couplings of this Hamiltonian, because
the couplings are changed in the course of the renormalization scheme,
and may differ from each other. We then replace the original four spins with
a Hamiltonian where only spins $1$ and $4$ remains, as detailed in
appendix B. In further iteration the remaining spins will combine
with their neighbors to form a new Eq. (\ref{h1234}).

\begin{figure}[ht!]
\centering
\includegraphics[width=85mm]{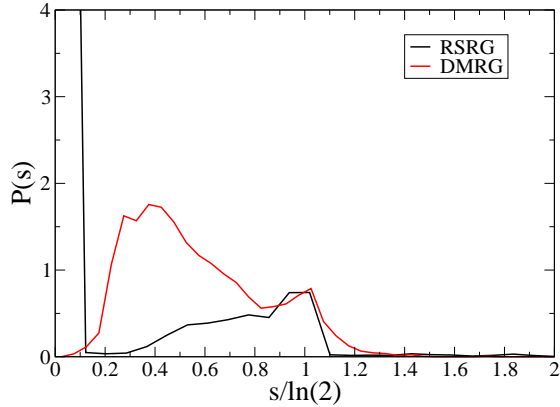}
\caption{Modified RSRG results (solid line) compared to DMRG results (dashed line) for $W=3.5,U=2.4$. The cluster size for the modified RSRG is 10000 for a system of 100 sites. Deviation from half-filling was considered large if it was larger than four. In the modified RSRG results there is a high peak at zero, which contains states of low entanglement that are not correctly described by the modified RSRG approximation, and a smaller peak around 1, in agreement with the DMRG results.  }
\label{rsrg2}
\end{figure}
\begin{figure}[ht!]
\centering
\includegraphics[width=60mm]{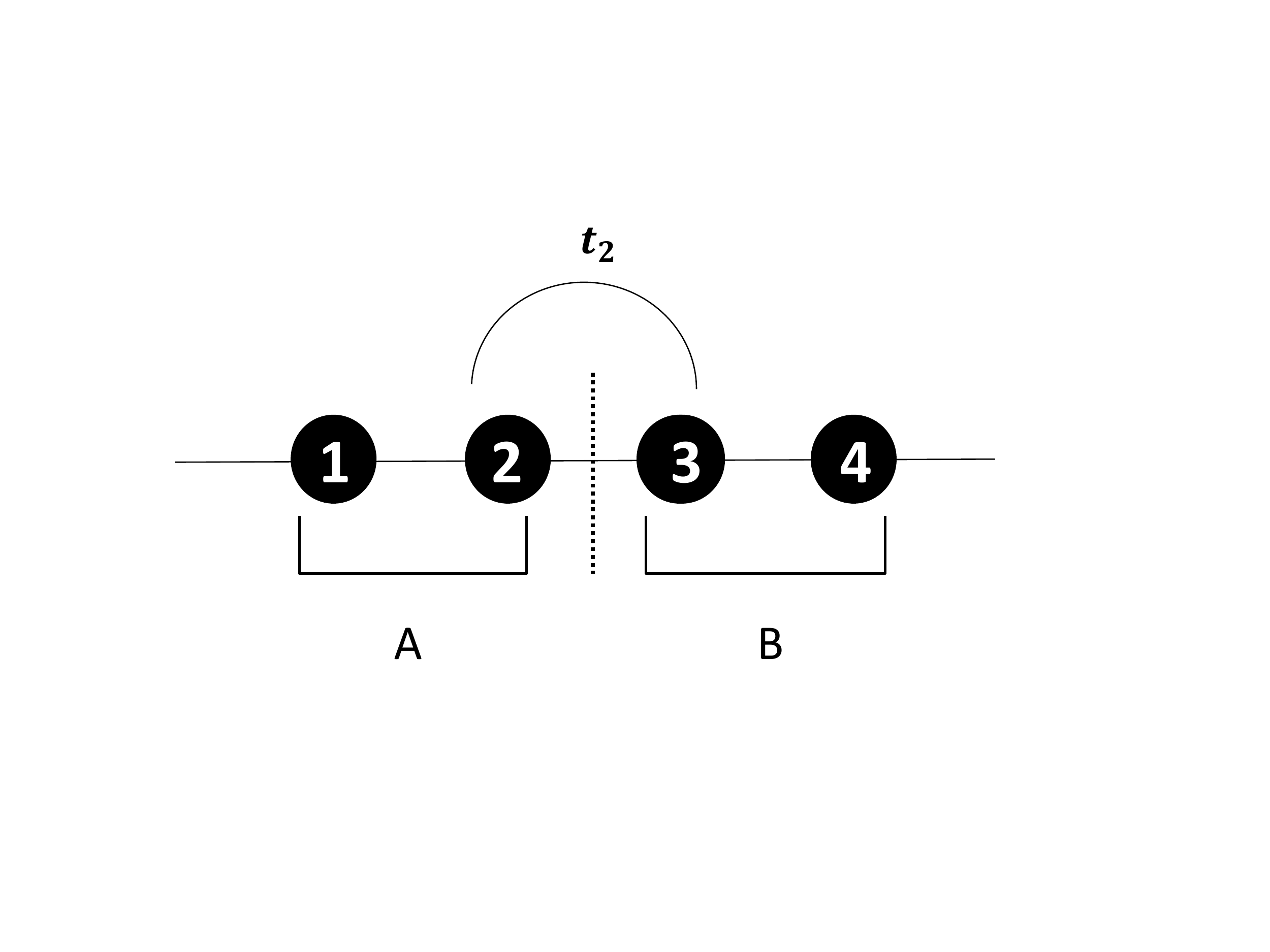}
\caption{A toy model with two fermions and $4$ sites described by the
  tight binding model (for the interacting cases- with nearest neighbors
  interaction, $U$). Inside each region (A,B) the hopping parameter
  is $t=1$. The hopping parameter between the regions $\tilde{t}$ is
  normally distributed in the range $(0.5,1.5)$.}
\label{model}
\end{figure}

This method has two main drawbacks. First, the renormalization procedure 
works best with long tail distributions, which is not the case for
the canonical Anderson model.
The reason is that perturbation theory might break down if next to the 
largest bond there is another bond of the same order of magnitude. 
Nevertheless, 
RSRG still works surprisingly well even for uniform distributions as 
can be seen in \cite{laf}. Second, we are considering a system with a fixed
filling (in particular half-filling). For the classical RSRG on the xx, 
xxx or Ising models, this is not a problem, since  singlets naturally 
preserve this condition. Nevertheless, the solution for the xxz system
contains also  other states 
(i.e., the $|\uparrow\uparrow\rangle $ and
$|\downarrow\downarrow\rangle $ states)  
which can break local and global half-filling (particle-hole) symmetry. In order to preserve 
the global particle-hole symmetry
(at least approximately) we modify the algorithm. here the 
first excited state may be chosen instead of the ground state in cases where
a large deviation from half-filling
is detected. Thus, fluctuations in the number of particles
are allowed, but are limited.
The entanglement entropy distribution obtained by this method is 
exhibited in Fig. \ref{rsrg2}.

 We can clearly see a high peak close to zero EE. This peak is
 related to the low entanglement peaks seen in the DMRG calculation (which
 is plotted for comparison in Fig. \ref{rsrg2}. This peak emerges from the
 flow structure of the RSRG, i.e., the renormalization group flow takes small
 values
 to zero.  Another peak at $\ln(2)$
 is also apparent. It seems that the modified RSRG, which is tailored for
 the strong disordered case, overestimates the influence of disorder on the
 peak centered at low values of entanglement and pushes it towards values
 close to zero. On  the other hand, the peak at 
 $\ln(2)$ related to a singlet traversing the boundary, is reproduced rather
 well.  The broadening of this peak reflects the site to site fluctuations in 
the values of $a_3$ and $b_3$, and hence the peak at $\ln(2)$ is broadened 
toward values lower than $\ln(2)$.
Larger values than $\ln(2)$ are related to several entangled pairs across 
the boundary. This also explains the absence of peaks at $n \ln(2)$, since
once the singlet entanglement across the boundary is broadened,
the probability of
several such singlets combining to an integer peak in the distribution becomes
rather negligible. This is the reason why only a single peak of at
$\ln(2)$ is seen both in DMRG as well as for modified RSGS.


\subsection{Conclusions}

The EE distribution of spinless fermions at half-filling in
the presence of repulsive interactions exhibits an interesting structure
of a broad peak at low entanglement values and a narrow one at $\ln(2)$.
DMRG calculations reveal this structure and the significance of the
disorder and the interaction strength. A modified RSRG method was
used to get a better understanding of this behavior. One peak is centered
at low values of EE and
shifts to even lower values as disorder (or interaction strength) is
enhanced.
A second peak appears around the EE value of $\ln(2)$. Unlike the case where
the Anderson model with only bond disorder is considered, this peak is
broadened and no additional peaks at higher integer multiplications of $\ln(2)$.
This is a result of the on-site disorder which breaks the symmetry across the
bond, and therefore no pure singlet states transverse the boundary and one
can not simply count the integer number of singlet states crossing the
the boundary. This leads to the difference between the strong disorder
behavior of the EE distribution between the bond-disordered case and the
on-site disordered one. Thus, even at extreme disorder the symmetry of the
system continues to play an important role in the behavior of the entanglement.

Specifically, the reason we see only a single peak is that the region of
parameters we investigate is an intermediate regime between the
antiferromagnetic regime and the RSP regime. In this range of parameters
we have clusters of antiferromagnetic spins. When the sub-region $L_A$ ends
up inside the antiferromagnetic cluster the entanglement will be very low,
since most spins are almost anti-parallel to each other.
However, the boundary of two clusters contains spins that are correlated.
If we cut the system at such a boundary, we can get a singlet. This is
the origin of the $\ln(2)$ peak.  If in addition there happens to be
some correlation between spins that are located far apart across the
boundary we can get higher EE. Such events are rare, but exist, and
are the origin of the right handed tail in the EE. Consequently,
no peak is created beyond $\ln(2)$.

The main difference between our model and other similar once is the on site
disorder in the fermionic representation or the local magnetic field in
the spin representation. This term encourages the creation of clusters,
since it compete the other parameters and ruins their domination.
Positive value will support the creation of an anti-ferromagnet while
negative values encourage the spins to align in parallel. The higher the
disorder is, the higher is the probability of a spin to yield to this term,
thus, the spins become less correlated, and the values of entanglement lower.
There are still enough values where the magnetic field is negligible
compared to the other terms and singlets emerge as in the RSP.
The presence of interactions also strengthens the anti-ferromagnetic
tendency, and also leads to lower entanglement values. T
his is summarized in table \ref{tb_phases}.

\begin{table*}[t]
\begin{center}
    \begin{tabular}{ |c|c|c|}
    \hline
 Model & Hamiltonian & Phase\\ 
&&\\ \hline\hline
XX model  & $\sum_j t_j(S_j^xS_j^x+S_j^yS_j^y)$ & Random singlet phase  \cite{das,fisher}\\
&&\\ \hline
XXZ model& $\sum_j t_j(S_j^xS_j^x+S_j^yS_j^y)+US_j^zS_j^z$ &  Anti-ferromagnet $(U>t)$ \cite{das,fisher}\\ 
&&\\\hline
Our model &$\sum_j t_j(S_j^xS_j^x+S_j^yS_j^y)+US_j^zS_j^z+\epsilon_j S_j^z$
 & Anti-ferromagnetic clusters\\&&\\\hline
    \end{tabular}
    \caption{ Although these spin models are similar, they give rise to
      different phases. In our model the magnetic field term competes
      with the other parameters and encourages the creation of anti-parallel
      spins while the other parameters push towards RSP, and the creation of
      singlet in the anti-ferromagnet range of parameters.}
    \label{tb_phases}
\end{center}
\end{table*}

The behavior in the presence of attractive interaction and in quasi-1d
dimensions are left for further study.

\acknowledgments

\section{APPENDIX A: TOY MODEL}

 The ground state energy of Hamiltonian (\ref{toyh}) is
\begin{equation}
\label{gs}
\small{
\begin{aligned}
 E_0= &\dfrac{1}{3}( - U - \dfrac{ 2^{2/3} T}{(R +\sqrt{-4 T^3 + R^2})^{1/3}} -  (\frac{R + \sqrt{-4 T^3 +R^2})}{2})^{1/3})\\
 &T=12 + 3 \tilde{t}^2 + U^2,  R=-18 U + 9 \tilde{t}^2 U + 2 U^3\\
   \end{aligned}}
\end{equation}
and the ground state eigenvector is

\be\small{
\ba
&\vec{V_0}=\frac{1}{N}(E_0(E_0 + U), 2 (E_0 + U), 2E_0,E_0( E_0 + U),E_0 \tilde{t}, E_0 \tilde{t}),\\
&N=\sqrt{2(2E_0 + (E_0\tilde{t})^2 + E_0^2(E_0 + U)^2 + (2 (E_0 + U)^2))}
\ea}
\label{vandn}
\ee

In the basis:

\be
\ba
&|1\rangle=a_1^+ b_3^+|0\rangle;
|2\rangle=a_1^+ b_4^+|0\rangle;
|3\rangle=a_2^+ b_3^+|0\rangle\\
&|4\rangle=a_2^+ b_4^+|0\rangle;
|5\rangle=a_1^+ a_2^+|0\rangle;
|6\rangle=b_3^+ b_4^+|0\rangle
\ea
\label{basis}
\ee

where $|0\rangle$ is the vacuum.

For a general real vector $\vec{v}=(v_1,v_2,..,v_6)$ in the toy model, the reduced density matrix over region A is given by

\be
 \rho_A= \left( \begin{array}{cccc}
v_1^2+v_2^2 & v_1v_3+v_2v_4 & 0 & 0   \\
v_1v_3+v_2v_4 & v_3^2+v_4^2 & 0 & 0   \\
0 & 0 & v_5^2 & 0 \\
0 & 0 & 0 & v_6^2 \\
\end{array} \right)
\label{rho}
\ee
Substituting the relevant values, and diagonalizing, we find the eigenvalues of $\rho_A$: 

\be
\ba
&\mu_1=\frac{1}{N^2}(E_0^4 + 4 E_0 U + 2 E_0^3 U + 2 U^2 + E_0^2 (4 + U^2)\\& - 2 \sqrt{E_0^4 + 2 E_0^3 U + (1 + E_0^2) U^2} |2 E_0 + U|)\\
 &\mu_2=\frac{1}{N^2}(E_0^4 + 4 E_0 U + 2 E_0^3 U + 2 U^2 + E_0^2 (4 + U^2)\\& + 2 \sqrt{E_0^4 + 2 E_0^3 U + (1 + E_0^2) U^2} |2 E_0 + U|)\\
 &\mu_3=\mu_4=\frac{t_2^2E_0^2}{N^2}
\ea 
\label{mu}
\ee

The entanglement entropy (EE) is given by
\be
 S=\sum_{i=1}^4 -\mu_i\log(\mu_i)
\ee
The entanglement distribution is obtained analytically via the relation
\be
P(S)=P(\tilde{t})|\frac{\partial{\tilde{t}}}{\partial{S}}|
\ee

For interacting cases ($U\neq0$), an exact analytic expression can be obtained,
but are too long to display. The first terms in the series expansion in
large $U$  of the exact expressions are a good approximation, as confirmed by
numeric calculations.

For negative $U$ values using the transformation $U=-1/X$, the large $U$ expansion is equivalent to the expansion around $X=0$: 

\be
\ba
&E_0\approx -\sqrt{2} - (1 + \frac{\tilde{t}^2}{2}) X+O(X^2)\\
&S\approx -\frac{\sqrt{3}}{4} \cosh^{-1}(7) + \log(4) +
 \frac{(-6 + \tilde{t}^2) \cosh^{-1}(7)}{8 \sqrt{6}}X+\\
&\frac{1}{576} (\sqrt{3} (276 + 12 \tilde{t}^2 + \tilde{t}^4)\cosh^{-1}(7)+  \\
   &-12 (36 - 36 \tilde{t}^2 + \tilde{t}^4 + 48 \tilde{t}^2 \log(|\tilde{t} X|)) X^2+O(X^3)\\
 &\frac{\partial{\tilde{t}}}{\partial{S}}\approx [((\tilde{t} X \cosh^{-1}(7))/(4 \sqrt{6}) + \\
 &\frac{1}{576} X^2 (\sqrt{3} (24 \tilde{t}+ 4 \tilde{t}^3) \cosh^{-1}(7)\\  
 &   -12 (-24\tilde{t} + 4 \tilde{t}^3 + 96 \tilde{t} \log(|\tilde{t} X|))))]^{-1}+O(X^2)
 \ea
\label{analytic} 
 \ee

\begin{figure}[ht!]
\centering
\includegraphics[width=90mm]{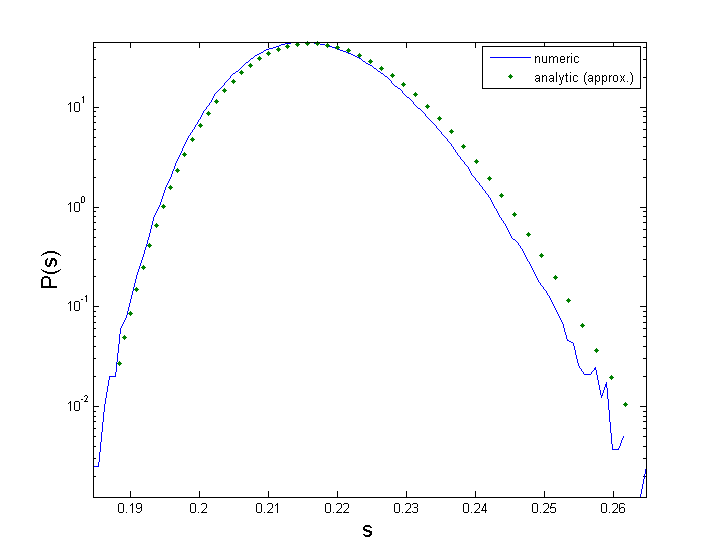}
\caption{EE semilog distribution for the toy model. The interaction strength is $U=-10$ (repulsive in the toy model convention) and the disorder distribution is normally distributed around 1. The solid line plot results from numeric diagonalization over $10^6$ realizations of the toy-model Hamiltonian (in Fock space) using MATLAB. Some numerical fluctuations appear at the bottom. The dotted lines are the analytic approximations discussed in the text.  The analytic approximated results (Eq.(\ref{analytic})) are a good approximation to the exact numeric ones. }
\label{ins1}
\end{figure}

Since the system is small and its parameters can be regarded as rising from the central limit theorem, we choose a Gaussian distribution for $\tilde{t}$. The variance of the Gaussian is a function of the disorder strength $W$.  The results are shown in Fig. (\ref{ins1}).

The toy model describes well the low entanglement values, but due to its small size, anti-ferromagnetic clusters do not have space to develop and hence no
$\ln(2)$ peak appears.
We used a combination of the toy model result and a Gaussian fit to
describe the DMRG results Fig. \ref{fig4}.

\section{Appendix  B: 
The Modified RSRG}

\begin{table*}[t]
\begin{center}
    \begin{tabular}{ |c||c|c|c|}
    \hline
 Ground state vector & $|v_1\rangle=|\uparrow\uparrow\rangle$ & $|v_2\rangle=|\downarrow\downarrow\rangle$ & $|v_3\rangle=a_3|\uparrow\downarrow\rangle +b_3|\downarrow\uparrow\rangle$ \\ 
&&&\\ \hline\hline
  1st order correction & $ \dfrac{U_1}{2}S_1^z+\dfrac{U_3}{2}S_4^z$ & $-\dfrac{U_1}{2}S_1^z-\dfrac{U_3}{2}S_4^z$ & $(\dfrac{U_1}{2}S_1^z-\dfrac{U_3}{2}S_4^z)(a_3^2-b_3^2)$. \\
&&&\\ \hline
2nd order correction &$C_1+t'(S_1^+S_4^-+S_1^-S_4^+)$ &  $C_2+t'(S_1^+S_4^-+S_1^-S_4^+)$ &  $C_3+t'(S_1^+S_4^-+S_1^-S_4^+)+U'S_1^zS_4^z$ \\ 
&&&\\\hline
    Couplings modification & $\epsilon_1\rightarrow\epsilon_1+U_1/2$  & $\epsilon_1\rightarrow\epsilon_1-U_1/2$
 & $\epsilon_1\rightarrow\epsilon_1+ U_1(a_3^2-b_3^2)/2$ \\ 
&$\epsilon_4\rightarrow\epsilon_4+U_3/2$&$\epsilon_4\rightarrow\epsilon_4-U_3/2$&$\epsilon_4\rightarrow\epsilon_4- U_3(a_3^2-b_3^2)/2$\\ \hline
Effective x-y coupling&$ 2t_1t_3(\dfrac{a_3b_3}{E_1-E_3}+\dfrac{a_4b_4}{E_1-E_4})$&$2t_1t_3(\dfrac{a_3b_3}{E_2-E_3}+\dfrac{a_4b_4}{E_2-E_4})$ &$ 2t_1t_3a_3b_3(\dfrac{1}{E_3-E_1}+\dfrac{1}{E_3-E_2})$\\ 
t'&&&\\\hline
Effective z coupling &$    0$&$    0$ & $-\dfrac{U_1U_3(a_3a_4-b_3b_4)^2}{2(E_3-E_4)}$\\
U'&&&\\\hline
Constant energy term &$\dfrac{t_1^2b_3^2+t_3^2a_3^2}{E_1-E_3}+\dfrac{t_1^2b_4^2+t_3^2a_4^2}{E_1-E_4}$&$\dfrac{t_1^2a_3^2+t_3^2b_3^2}{E_2-E_3}+\dfrac{t_1^2a_4^2+t_3^2b_4^2}{E_2-E_4}$&$\dfrac{t_1^2b_3^2+t_3^2a_3^2}{E_3-E_2}+\dfrac{t_1^2a_3^2+t_3^2b_3^2}{E_3-E_2}$
\\$C_i, i=1..3$&&&$+\dfrac{(U_1^2+U_2^2)(a_3a_4-b_3b_4)^2}{4(E_3-E_4)}$
\\
 \hline
    \end{tabular}
    \caption{ Perturbation theory results for one RSRG iteration. For each ground state the coupling is modified according to the original values of the couplings, the states' energies and the wave function coefficients. The wavefunction coefficients are assumed to be real for convenience and $C_i, i=1,2,3$ are non universal constant energy terms }
    \label{tb1}
\end{center}
\end{table*}

We use second order perturbation theory to describe the effect of the sites 2-3 on their neighbors. Each of the possible eigenstates change the Hamiltonian couplings in a different manner. Thus, the renormalized Hamiltonian is not the same for all the cases, and it depend also on the couplings' values.

For all the possible g.s. vectors the first order correction modifies the random magnetic field coefficient $\epsilon$ of spins 1 and 4, while the second order correction give rise to the effective couplings between them. For all cases the spins couple in the x-y plane. For $|v_3\rangle$ there is  also a coupling in the z direction.

Table \ref{tb1} shows the perturbation terms and the modifications in the couplings.

\end{document}